\documentclass[twocolumn,showkeys,pra]{revtex4}
\usepackage{epsf}
\usepackage{amsmath,amsthm,amssymb}
\usepackage{color}
\usepackage{dcolumn}
\usepackage{bm}
\usepackage{graphicx,epsfig}
\usepackage[dvipsnames]{xcolor}

\graphicspath{{fig/}}
\everymath{\displaystyle}

\begin{document} 

\title{Foldy-Wouthuysen transformation and multiwave states of a graphene electron in external fields and free (2+1)-space}

\author{Alexander J. Silenko$^{1,2,3}$}
\email{alsilenko@mail.ru}

\affiliation{$^1$Bogoliubov Laboratory of Theoretical Physics, Joint
Institute for Nuclear Research, Dubna 141980, Russia}
\affiliation{$^2$Institute of Modern Physics, Chinese Academy of
Sciences, Lanzhou 730000, China}
\affiliation{$^3$Research Institute for Nuclear Problems, Belarusian
State University, Minsk 220030, Belarus}

\date{\today}

\begin{abstract}
The relativistic Foldy-Wouthuysen transformation is used for an advanced description of planar graphene electrons in external fields and free (2+1)-space. It is shown that the initial Dirac equation should {by based on the usual $(4\times4)$} Dirac matrices but not {on the reduction of matrix dimensions and the use of $(2\times2)$} Pauli matrices. {The latter approach does not agree with the experiment.} The spin of graphene electrons is not the {one-value} spin and takes the values $\pm1/2$. The exact Foldy-Wouthuysen Hamiltonian of a graphene electron in uniform and nonuniform magnetic fields is derived. The exact energy spectrum agreeing with the experiment and exact Foldy-Wouthuysen wave eigenfunctions are obtained. These eigenfunctions describe multiwave (structured) states in (2+1)-space. It is proven that the Hermite-Gauss beams exist even in the free space. In the multiwave Hermite-Gauss states, graphene electrons acquire nonzero effective masses dependent on a quantum number and move with group velocities which are less than the Fermi velocity. Graphene electrons in a static electric field also can exist in the multiwave Hermite-Gauss states defining non-spreading coherent beams. These beams can be accelerated and decelerated.
\end{abstract}


\maketitle








\section{Introduction}

The graphene is an allotrope of carbon consisting of a single layer of atoms arranged in a hexagonal lattice nanostructure \cite{GeimNov}. It possesses many extraordinary properties. In particular, current carriers in the graphene (electrons) are massless. Since the graphene has a 2-dimensional structure, graphene electrons exist in (2+1)-space. Certainly, they are described by the Dirac equation in this space. A solution of this equation defines all fundamental properties of the graphene electron. In the present study, we consider the planar graphene electron.

Unlike \emph{all} previous studies of the graphene electron, we use the relativistic Foldy-Wouthuysen (FW) transformation of the initial Dirac equation developed in Refs.
\cite{JMP,PRA2015}. As a result, we fulfill the exact description of the graphene electron in a magnetic field and the high-precision description in an electric field. The relativistic FW transformation can also be used for a particle in nonstationary fields \cite{PRAnonstat}. The FW representation in relativistic quantum mechanics (QM) {is equivalent to} the Schr\"{o}dinger {representation in nonrelativistic QM and operators in these representations (but not in the Dirac one) are quantum-mechanical counterparts of corresponding classical variables (see Ref. \cite{PRAFW})}. The solution for the magnetic field is similar to the Landau result \cite{Landau,LL3} and gives one the exact FW wave eigenfunctions and the exact energy spectrum agreeing with experimental data \cite{GrapheneMiller,GraphenePCheng,Jiang}. 

An analysis of paraxial multiwave states of graphene electrons in the free (2+1)-space and a static uniform electric field demonstrates the existence of Hermite-Gauss (HG) states and the possibility to accelerate and decelerate the electrons without a lost of coherence.
We also describe extraordinary 
properties of graphene electrons in the HG states, namely, acquiring nonzero effective masses dependent on a quantum number and moving with group velocities which are less than the carrier velocity. The latter velocity is equal to the Fermi one $v_F\approx1 \times 10^6$ m/s.

The classical description of spin in (2+1)-space fulfilled in Sec. \ref{classical} shows that the spin is a real and important physical parameter. We consider relativistic FW transformations in Sec. \ref{DiffGrav}. The exact solutions for a graphene electron in a magnetic field are obtained in Sec. \ref{stanbox}. In Secs. \ref{general} and \ref{Electric}, we analyze multiwave (structured) HG states and extraordinary 
properties of graphene electrons in the free space and a static electric field. Section \ref{Discussionsummary} summarizes the results obtained.

We assume that $\hbar=1,~c=1$ but include $\hbar$ and
$c$ into some formulas when this inclusion clarifies the problem. In the considered case, the role of $c$ plays the Fermi velocity $v_F$. {As a result, we usually suppose that $v_F=1$.} While the graphene electrons are massless, we take into account the existence of massive chiral quasiparticles in the bilayer graphene (see Ref. \cite{GrapheneGeim} and references therein) and keep the mass in many formulas. 

\section{Classical description of spin in (2+1)-space}\label{classical}

There is a wonderful similarity between classical and quantum-mechanical Hamiltonians and equations of spin motion in (3+1)-space. This similarity becomes evident when the FW representation \cite{FW} is used (see Ref. \cite{PRAFW} and references therein). It is very natural to suppose that such a similarity remains valid in (2+1)-space. Therefore, a classical analysis of spin in (2+1)-space plays an important role.

In this case, the antisymmetric tensor of electromagnetic field 
takes the form ($\mu,\nu=0,1,2$) 
\begin{equation} F_{\mu\nu}=\partial_\mu A_\nu-\partial_\nu A_\mu=\left(\begin{array}{ccc} 0 & E_1 & E_2 \\ -E_1 & 0 & -B \\ -E_2 & B & 0 \end{array}\right), \label{ast} \end{equation} where $A_\mu=(A_0,A_1,A_2)=(A_0,-\bm A)$ is the field potential and $\bm E=(E_1,E_2)$ and $B$ are the electric and magnetic fields. The difference between upper and lower indices is only for the field potential ($A^1=-A_1,\,A^2=-A_2$).

Vector products of two components of any vectors correspond to third components in the (3+1)-dimensional space: $\bm C\times\bm D=C_1D_2-C_2D_1$. Certainly, this \emph{scalar} quantity can have any sign. To avoid any misleading, we will often write such a vector product in the form $(\bm C\times\bm D)_3$ demonstrating that this is a scalar.

{Taking into account the Lorentz transformations of electromagnetic fields also shows that $B$ can have any sign.}

The (2+1)-OAM tensor is given by
\begin{equation} {\cal L}^{\mu\nu}=\sum{(x^{\mu}p^\nu-x^{\nu}p^\mu)}. \label{OAML} \end{equation} The spatial components of this tensor read
\begin{equation} {\cal L}^{12}=-{\cal L}^{21}=L,
\label{OAMv} \end{equation} where $L$ is the OAM which can have any sign and can be equal to zero. The summation takes place for an ensemble of particles.

In classical physics and QM {in (3+1)-space}, the \emph{conventional} spin is defined in the particle rest frame. The total angular momentum is the sum of the orbital angular momentum (OAM) and the rest-frame spin. While the total angular momentum can also be defined as the sum of the OAM and the laboratory-frame spin, this definition needs the use of noncommutative geometry (see Ref. \cite{PRAFW} and references therein).

The conventional classical (rest-frame) spin {in (3+1)-space} is defined as the internal OAM of a compound particle in its rest frame. {The same definition of spin can be used in (2+1)-space. In this case,} the classical spin depends on a hidden internal motion in the compound particle. The OAM relative to any axis is conditioned by the distance to this axis and the momentum of compound particle as a whole. For the compound particle, there is no summation in Eq. (\ref{OAML}).

Since the OAM can have any sign or be equal to zero, the spin also has this property. As a result, the spin in QM of a Dirac particle can be equal to $\pm1/2$. Therefore, the spin coupling with the magnetic field, $sB$, can be nonzero and can has any sign even for a fixed sign of $B$. QM in the FW representation always agrees with classical physics and this agreement has rigorous substantiation \cite{PRAFW}. Classical and quantum-mechanical equations of spin motion are similar. The spin is an additional degree of freedom. The main distinguishing feature of spin in QM is its quantization. 

{All these spin properties take place when the spin is defined as the internal OAM in the particle rest frame. However, mathematical physics states that the spin in (2+1)-space has different meaning. The quantum-mechanical analysis shows \cite{Graphenesimulations} that the spin of a specific particle can have only one value, positive or negative. Signs of the spin differ for particles and antiparticles \cite{Graphenesimulations}. Such a spin plays the role of the flavor in high energy physics \cite{Graphenesimulations} or can be regarded as the pseudospin when it indicates the sublattice \cite{GrapheneGeim}.}

{Nevertheless, this situation can be realized only for a particle in a pure (2+1)-space. In contrast, any graphene sheet is immersed in the (3+1)-space and interacts with photons and external fields in this space. In the latter case, the (pseudo)vector of total angular momentum should be conserved. It will be shown in the next section that the one-value graphene spin is inappropriate when a graphene sheet is considered in the (3+1) laboratory frame.}

The analysis fulfilled shows that the spin in the (2+1)-space immersed in the (3+1)-space is a real and important physical parameter which cannot be reduced to the one-value spin.

\section{Initial Dirac equation and relativistic Foldy-Wouthuysen transformation in (2+1)-space}\label{DiffGrav}

The relativistic FW transformations in (2+1)- and (3+1)-spaces are identical. Therefore, we use in the present study the relativistic method first developed in Ref. \cite{JMP} and presented in the final form in Ref. \cite{PRA2015}. In the general case, this method uses an expansion of the final result in a series in increasing powers of $\hbar$ and gives one \emph{exact} expressions for leading terms in the FW Hamiltonian proportional to the
zero and first powers of the Planck constant and for such terms proportional to $\hbar^2$ which describe contact interactions. 

A great advantage of the FW representation \cite{FW}
is the simple form of operators corresponding to classical
observables. In this representation, the Hamiltonian and all
operators are even, i.e., block-diagonal (diagonal in two
spinors). The passage to the classical limit usually reduces to a
replacement of operators in quantum-mechanical Hamiltonians
and equations of motion with the corresponding classical
quantities. The possibility of such a replacement, explicitly or
implicitly used in practically all works devoted to the FW
transformation, has been rigorously proved for the stationary case
in Ref. \cite{JINRLett12}. 

The initial Hamiltonian operator 
can be split into even and odd operators commuting and anticommuting with the Dirac operator
$\beta$, respectively:
\begin{equation} {\cal H}=\beta{\cal M}+{\cal E}+{\cal
O},~~~\beta{\cal M}={\cal M}\beta, ~~~\beta{\cal E}={\cal E}\beta,
~~~\beta{\cal O}=-{\cal O}\beta. \label{eq3} \end{equation} The
even operators ${\cal M}$ and ${\cal E}$ and the odd operator
${\cal O}$ are diagonal and off-diagonal in two spinors,
respectively. For a
Dirac particle, the ${\cal M}$ operator usually reduces to the
particle rest energy $mc^2$:
\begin{equation} {\cal H}_D=\beta mc^2+{\cal E}+{\cal
O}. \label{eq3Dirac} \end{equation}
{When
\begin{equation} [{\cal M},{\cal O}]=0,\qquad [{\cal E},{\cal O}]=0,\qquad [{\cal M},{\cal E}]=0 \label{exact} \end{equation} and the initial Hamiltonian is independent of time, the relativistic FW transformation is exact \cite{JMP,PRA2015,PRA2008}. In this case, the operators of the FW transformation and the inverse transformation read
\begin{equation}\begin{array}{c}
U_{FW}=\frac{\epsilon+{\cal M}+\beta{\cal O}}{\sqrt{2\epsilon(\epsilon+{\cal M})}},\qquad
U_{FW}^{-1}=\frac{\epsilon+{\cal M}-\beta{\cal O}}{\sqrt{2\epsilon(\epsilon+{\cal M})}},\\ \epsilon=\sqrt{{\cal M}^2+{\cal O}^2}.
\label{eq18} \end{array} \end{equation}
We should mention that the operator $\sqrt{\epsilon}$ is even and diagonal and is defined by $$\sqrt{\epsilon}\equiv\sqrt{{\rm diag}(\epsilon_{ii})}={\rm diag}\left(\sqrt{\epsilon_{ii}}\right).$$
The exact FW Hamiltonian is given by
\begin{equation}
{\cal H}_{FW}=\beta \epsilon+{\cal E}.
\label{eq17}
\end{equation}}

{When the commutation relations (\ref{exact}) are not satisfied, the transformation operator $U_{FW}$ can also be used, but the obtained \emph{approximate} relativistic FW Hamiltonian is different \cite{JMP,PRA2015,PRA2008}.}
In the general case, it has the form \cite{PRA2015,PRA2008}
\begin{equation}\begin{array}{c}
{\cal H}_{FW}=\beta\epsilon+ {\cal E}+\frac 14\Biggl\{\frac{1}
{2\epsilon^2+\{\epsilon,{\cal M}\}},\biggl(\beta\left[{\cal O},[{\cal O},{\cal
M}]\right]\\-[{\cal O},[{\cal O},{\cal
F}]]\biggr)\Biggr\},\quad \epsilon=\sqrt{{\cal M}^2+{\cal
O}^2},\quad {\cal F}={\cal E}-i\hbar\frac{\partial}{\partial t}.
\end{array} \label{MHamf} \end{equation} In this equation, external fields can be
nonstationary.

The (2+1) Dirac equation has the same form as the usual one but the space is reduced:
\begin{equation} i\gamma^\mu(\partial_\mu+ieA_\mu)\psi=m\psi, \qquad \mu=0,1,2. \label{inDirac} \end{equation} For graphene electron, $m=0$. {The Dirac matrices satisfy the usual relation}
$$\gamma^\mu\gamma^\nu+\gamma^\nu\gamma^\mu=2g^{\mu\nu},\qquad g^{\mu\nu}={\rm diag}
\{1,-1,-1\},$$ {where $g^{\mu\nu}$ is the metric tensor in the (2+1) Minkowski space.}
All vectors have two components, e.g., $\bm p=-i\hbar\left(\frac{\partial}{\partial x^1},\frac{\partial}{\partial x^2}\right)$, and $p_0={\cal H}$.
The Dirac equation (\ref{inDirac}) takes the form
\begin{equation} [\gamma^0(p_0-eA_0)-\bm\gamma\cdot\bm\pi+m]\psi=0,\qquad \bm\pi=\bm p-e\bm A, \label{tr3Dirac} \end{equation} where $\bm\pi$ and $\bm p$ are the kinetic and canonical (generalized)
momenta. As a result, the initial equation for the Dirac Hamiltonian reads
\begin{equation} {\cal H}_D\psi=(\bm\alpha\cdot\bm\pi+\beta m+eA_0)\psi,\quad \bm\alpha=\gamma^0\bm\gamma,\quad \beta\equiv\gamma^0. \label{Ha3Dirac} \end{equation}
For this equation, $$ {\cal E}=eA_0,\qquad {\cal O}=\bm\alpha\cdot\bm\pi. $$ For electrons, $e=-|e|$.

There are different forms of the Dirac matrices in (2+1)-space. The conventional approach used, e.g., in gravity (see Refs. \cite{Unal,Koke}) and electromagnetism \cite{RJackiw} consists in the reduction of the matrix {dimensions}: \begin{equation}\beta=\gamma^0=\sigma_3,\qquad\gamma^1=i\sigma_1,
\qquad\gamma^2=i\sigma_2,\label{reduction} \end{equation} where $\sigma_i~(i=1,2,3)$ are the $2\times2$ Pauli matrices. However, the following analysis shows 
that this reduction {is unnecessary and inappropriate for the considered problem}. 

It is sufficient to consider the 
planar graphene electron in a magnetic field. For the reduced spin matrices (\ref{reduction}), the {exact} relativistic FW Hamiltonian takes the form 
\begin{equation}\begin{array}{c}
{\cal H}_{FW}=\sigma_3\sqrt{m^2+\bm\pi^2-e\sigma_3B}.
\end{array} \label{Hamfu} \end{equation}
Since this Hamiltonian is diagonal, states with positive and negative total energies are separated. Lower and upper components of the two-component wave function vanish for positive-energy and negative-energy states, respectively. As a result, Eq. (\ref{Hamfu}) can be presented as follows:
\begin{equation}\begin{array}{c}
{\cal H}_{FW}=\left(\begin{array}{cc} \sqrt{m^2+\bm\pi^2-eB} & 0 \\ 0 & 
-\sqrt{m^2+\bm\pi^2+eB} \end{array}\right).
\end{array} \label{Hamfd} \end{equation}

Equations (\ref{Hamfu}) and (\ref{Hamfd}) clearly show that the Dirac equation with the reduced matrices (\ref{reduction}) does not describe the true spin. It follows from these equations that, in particular, a spin-1/2 particle in positive-energy states has \emph{only one} spin value. The spin characterized by Eqs. (\ref{Hamfu}) and (\ref{Hamfd}) is not an additional
degree of freedom. {We can show that these properties contradict to a consideration of the graphene sheet in the (3+1) laboratory frame. In this frame, (2+1) OAM becomes a (pseudo)vector normal to the graphene sheet and contributes to the total angular momentum. However, the use of the one-value (2+1) spin meets insurmountable difficulties. It is well know that any massless particle in the (3+1)-space has two helicity states ($h=\bm s\cdot\bm p/p=\pm1/2$ for a Dirac fermion). Therefore, the spin of any massless particle has \emph{two} basic states. Certainly, this conclusion covers (2+1) graphene electrons in the (3+1) laboratory frame. While the above-mentioned states describe the polarization in the plane of the sheet, in the (3+1)-space appropriate coherent superpositions of them define two other independent states characterizing two opposite spin directions orthogonal to the sheet.} 

{This consideration demonstrates that the (2+1) graphene electron in the (3+1) laboratory always has \emph{two} independent spin states with $s=\pm1/2$. Equations (\ref{Hamfu}) and (\ref{Hamfd}) show that the spin contributes to the energy of the particle and this contribution is important. As a result, we can conclude that the conventional approach consisting in the reduction of the matrix dimensions is inapplicable to graphene electrons.}

{Instead,} one can use the original $4\times4$ Dirac matrices \cite{Lozovik}. We apply this approach in the present study. When the original Dirac matrices are utilized, the relativistic FW Hamiltonian has the form
\begin{equation}\begin{array}{c}
{\cal H}_{FW}=\beta\epsilon+ eA_0+\frac e8\biggl\{\frac{1}
{\epsilon(\epsilon+m)},\Sigma_3(\bm\pi\times\bm E\\-\bm E\times\bm\pi)_3\biggr\},\qquad \epsilon=\sqrt{m^2+\bm\pi^2-e\Sigma_3B}.
\end{array} \label{THamf} \end{equation}
In this equation, only the field potentials and their first derivatives are taken into account. The applied method gives \emph{exact} expressions for such terms \cite{PRA2015}. Here and below, we use the standard form of the Dirac matrices \cite{BLP}. The total energy can be positive and negative.

Equation (\ref{THamf}) explicitly shows that the rudimentary spin in (2+1)-space is an important characteristic of the particle polarization but not an {one-value (pseudo)spin}. The eigenvalues $\mathfrak{S}_3=\pm1$ of the operator $\Sigma_3$ are scalar values but not projections of any vector. The spin is an additional degree of freedom and its value can be positive or negative. Thus, the use of the original $4\times4$ Dirac matrices conserves the agreement between QM in {(2+1)- and (3+1)-spaces}. This fundamental conclusion can be extended on Dirac particles in (2+1)-gravity, {when the (2+1)-space originates from the (3+1)-space \cite{PhysRevD2014}.}

We should also mention that spin interactions of anyons with an electromagnetic field have been considered in Ref. \cite{Horvathy}.

\section{Graphene electron in electric and magnetic 
fields}\label{stanbox}
\subsection{Exact solutions for a graphene electron in a magnetic 
field}\label{Exact}

When the original $4\times4$ Dirac matrices are used, the FW Hamiltonian for a graphene electron in a static magnetic field is exact and reads 
\begin{equation}\begin{array}{c}
{\cal H}_{FW}=\beta\sqrt{m^2+\bm\pi^2-e\Sigma_3B}.
\end{array} \label{eHamf} \end{equation} We underline that the magnetic field can be nonuniform. A similar situation takes place for a Dirac particle \cite{JMP,Case,Energy1,Energy3} and a spin-1 particle with the normal ($g=2$) magnetic moment \cite{spin1} in (3+1)-space. The corresponding FW Hamiltonians are also exact. The similarity of the FW Hamiltonians in (2+1)- and (3+1)-spaces substantially simplifies finding the exact energy spectrum and exact wave eigenfunctions for the Hamiltonian (\ref{eHamf}). 

 Fortunately, exact solutions for a graphene electron ($m=0,~e=-|e|$) in a uniform magnetic field can be obtained from the Landau solutions for (3+1)-space. We suppose that $B>0$. There are two appropriate gauges \cite{LL3}. For the symmetric gauge $A_\phi=Br/2,\,A_r=0$, the squared Eq. (\ref{eHamf}) takes the form
\begin{equation}
\begin{array}{c}
-\nabla^2+ieB\frac{\partial}{\partial\phi}+\frac{e^2B^2r^2}{4}-2es_zB=\mathfrak{E}^2,
\\ \nabla^2=
\frac{\partial^2}{\partial r^2}+\frac1r\frac{\partial}{\partial
r}+\frac{1}{r^2}\frac{\partial^2}{\partial\phi^2},
\end{array}
\label{eqpqmef}
\end{equation} where $\mathfrak{E}$ is the energy.

The energy spectrum is defined by the well-known eigenvalues of the operator $\bm\pi^2$ \cite{LL3} and is gived by ($e=-|e|$)
\begin{equation}
\begin{array}{c}
\mathfrak{E}=\pm\sqrt{\left(2n+1+|\ell|+\ell+2s_z\right)|e|B}.
\end{array}
\label{botmt}
\end{equation}
Here $s_z=\pm1/2$, $n=0,1,2,\dots$, and $\ell$ (more precisely, $\hbar\ell$) is the OAM which is a scalar (but not a pseudovector) in (2+1)-space. {$s_z$ is an expectation value of the corresponding spin operator ($+1/2$ or $-1/2$) and} $\ell$ is integer. The positive and negative signs define states with positive and negative total energies (graphene electrons and holes, respectively). The corresponding wave eigenfunctions are also defined by the Landau solution for the symmetric gauge \cite{LL3} (see also Refs. \cite{Energy3,paraxialLandau}). They are based on the Laguerre polynomials and are given by
\begin{equation}
\begin{array}{c}
\Phi_{FW}={\cal A}\exp{(i\ell\phi)},\qquad \int{\Phi_{FW}^\dag\Phi_{FW} rdrd\phi}=1,\\
{\cal A}=\frac{C_{n\ell}}{w_m}\left(\frac{\sqrt2r}{w_m}\right)^{|\ell|}
L_n^{|\ell|}\left(\frac{2r^2}{w^2_m}\right)\exp{\left(-\frac{r^2}{w^2_m}\right)}\eta,\\
C_{n\ell}=\sqrt{\frac{2n!}{\pi(n+|\ell|)!}},\qquad
w_m=\frac{2}{\sqrt{|e|B}},
\end{array}
\label{Lenergy}
\end{equation} where the real function ${\cal A}$ defines the amplitude of the beam, and $L_n^{|\ell|}$ is the generalized Laguerre polynomial. The wave function $\Phi_{FW}$ is the upper spinor for the FW bispinor wave function in positive-energy states: $\Psi_{FW}=\left(\begin{array}{c} \Phi_{FW} \\ 0 \end{array}\right)$. If negative-energy states are disregarded, the spin function $\eta$ is an eigenfunction of the Dirac operator $\Sigma_z$ (cf. Ref. \cite{Energy3}):
$$\Sigma_z\eta^\pm=\pm\eta^\pm,\quad \eta^+=\left(\begin{array}{c} 1 \\ 0 \\ 0 \\ 0 \end{array}\right),\quad \eta^-=\left(\begin{array}{c} 0 \\ 1 \\ 0 \\ 0 \end{array}\right).$$

Another gauge is also possible \cite{Landau,LL3}: $A_x=-By,\,A_y=0$. For this gauge, the energy eigenvalues can be obtained from the following equation:
\begin{equation}
\begin{array}{c}
-\nabla^2+2ieBy\frac{\partial}{\partial x}+e^2B^2y^2-2es_zB=\mathfrak{E}^2,
\\ \nabla^2=
\frac{\partial^2}{\partial x^2}+\frac{\partial^2}{\partial y^2}.
\end{array}
\label{eqpqmeL}
\end{equation}
The energy spectrum is given by
\begin{equation}
\begin{array}{c}
\mathfrak{E}=\pm\sqrt{\left(2N+1+2s_z\right)|e|B},
\end{array}
\label{botmL}
\end{equation} where $N=0,1,2,\dots$. Evidently, Eqs. (\ref{botmt}) and (\ref{botmL}) define the same set of energy levels. We underline the importance of the spin. {The energy levels (\ref{botmt}) and (\ref{botmL}) can be presented as follows:
\begin{equation}
\mathfrak{E}=\pm\sqrt{2\mathfrak{n}|e|B},\qquad \mathfrak{n}=0,1,2,\dots
\label{botmM}
\end{equation}}

{The unit in Eqs. (\ref{botmt}) and (\ref{botmL}) is caused by the oscillator-like form of initial equations for $\mathfrak{E}^2$. The energy of the lowest level of oscillator is positive. Therefore, the zero energy level is occupied only when $s_z=-1/2$ and is nondegenerate. All other energy levels are degenerate and can be occupied by electrons with any (positive or negative) spin. The same situation takes place for massive electrons in (3+1)-space (see relativistic formulas in Refs. \cite{Energy2,Rajabi,paraxialLandau}). For positively charged particles, as follows from Eq. (\ref{eHamf}), the zero energy level is occupied when $s_z=+1/2$.}

{The Zeeman splitting for graphene electrons significantly differs from that for atoms. As follows from Eqs. (\ref{botmt}), (\ref{botmL}), and (\ref{botmM}), the energies of Zeeman levels are proportional to $\sqrt{B}$. Magnetic moments of graphene electrons cannot be introduced because these energies cannot be presented in the form $\mu B$. Levels with the same quantum numbers $n,\,l$ but opposite signs of spin are neighbouring.}

Equations (\ref{botmt}), (\ref{botmL}), and (\ref{botmM}) perfectly agree with experimental data \cite{GrapheneMiller,GraphenePCheng,Jiang}. In fact, all energy levels are shifted by the value of the Dirac energy $E_D$ \cite{GraphenePCheng}. {The equation  (\ref{botmL}) is a final result in all theoretical papers (see Refs. \cite{CastroNeto,GrapheneKuru} and references therein). Nevertheless, derivations in these papers were based on the use of the \emph{reduced} Dirac matrices (\ref{reduction}) in initial Eq. (\ref{Ha3Dirac}). One of preferences of the FW transformation is a clear interpretation of obtained results \cite{JMP,PRAFW}. Equations (\ref{Hamfu}) and (\ref{Hamfd}) which are easily obtained with this transformation unambiguously show that the energy spectrum of electrons ($e=-|e|$) does
not contain the positive-energy level $\mathfrak{E}=\sqrt{2|e|B},~\mathfrak{n}=1$. In this case, there is a gap between positive-energy levels with $\mathfrak{n}=0$ and $\mathfrak{n}=2$. Therefore, the quantum-mechanical approach based on the reduction of the matrix dimensions disagrees with experimental data.}

{In all previous studies, supersymmetric QM was used. The validity of this approach has been confirmed in many papers (see, e.g., Refs. \cite{RJackiw,Haldane,GrapheneKuru}). Supersymmetric nature of the (2+1) Dirac Hamiltonian in the presence of electric and magnetic fields has been reported
in Refs. \cite{GrapheneKuru,GraphenePeres,CastilloCeleita,GrapheneNath}.} {If supersymmetric QM is appropriately used, it also leads to correct results for the considered problem. In our notation, the squared supersymmetric Hamiltonian obtained in Ref. \cite{RJackiw} reads
\begin{equation}\begin{array}{c}
{\cal H}^2={\bm\pi^2-e\sigma_3B}.
\end{array} \label{Hamqd} \end{equation}
Here the charge $e$ missed in Ref. \cite{RJackiw} is added and $m=0$. The extraction of the square root results in \begin{equation}{\cal H}=\sigma_3\sqrt{\bm\pi^2-e\sigma_3B}.\label{Hamnb} \end{equation}
The necessity of the matrix $\sigma_3$ outside the square root follows from the Hamiltonian for free particles. Evidently, Eq. (\ref{Hamnb}) is equivalent to Eq. (\ref{Hamfu}) for $m=0$. However, Eq. (\ref{Hamnb}) was not written down in precedent papers. We suppose that the confusion with the energy spectrum has happened due to the emphasis on the zero energy level and lack of attention to other energy levels in the literature.}

{We can conclude that the generally accepted energy spectrum for a graphene electron in a uniform magnetic field corresponds to the initial equation with the usual $(4\times4)$ Dirac matrices and is incompatible with its conventional replacement by the Pauli ones.} {While supersymmetric QM is correct, the FW transformation method allows one to obtain needed results more straigthforwardly and easily. A distinctive feature of the FW representation is its equivalence to the Schr\"{o}dinger one in nonrelativistic QM. FW wave functions have the probabilistic interpretation and operators in the FW representation are counterparts of corresponding classical variables \cite{PRAFW}.} 

The eigenfunctions of Eq. (\ref{eqpqmeL}) can be simply taken from Refs. \cite{Landau,LL3} (with the rejection of $z$ coordinate and the addition of the spin wave function):
\begin{equation}
\begin{array}{c}
\Phi_{FW}=\exp{(ip_xx)}\psi(y)\eta,\qquad \int{\psi^\dag\psi dy}=1,\\
\psi(y)=C_{N}\exp{\left[-\frac{2(y-y_0)^2}{w_m^2}\right]}H_N\left[\frac{2(y-y_0)}{w_m}\right],\\
C_{N}=\frac{1}{\pi^{1/4}w_m^{1/2}\sqrt{2^{N-1}N!}},\qquad y_0=-\frac{p_x}{eB},
\end{array}
\label{LenergL}
\end{equation} where $H_N$ is the Hermite polynomial. 
We note that the exact solutions (\ref{Lenergy}) and (\ref{LenergL}) define multiwave states formed by infinite continuums of partial de Broglie waves but not by single de Broglie waves having definite momenta. 

\subsection{Graphene electron in crossed electric and magnetic 
fields}\label{twofields}

An important problem is a description of a graphene electron in crossed uniform electric and magnetic fields. In this case, the contraction or collapse of Landau levels takes place \cite{GraphenePeres,CastilloCeleita,GrapheneNath,VLukose,Ma,Betancur,Ates}. Similar effects occur in a uniform electric field and a pseudo-mangetic field induced by a strain \cite{Phan,Le,LeLe} and in uniform magnetic and radial electric fields \cite{Nimyi}.

In the present study, we restrict ourselves to a discussion of quantum-mechanical equations of motion in the FW representation. We consider a comprehensive investigation of the problem as an outlook. 

In the analyzed case, $A_0=Ey$ and $\bm E=E\bm e_y$. The FW Hamiltonian (\ref{THamf})  takes the form
\begin{equation}\begin{array}{c}
{\cal H}_{FW}=\beta\epsilon+ eEy+\frac {ev_F}{8}\biggl\{\frac{1}
{\epsilon^2},\Sigma_3(\bm\pi\times\bm E\\-\bm E\times\bm\pi)_3\biggr\},\qquad \epsilon=\sqrt{v_F^2\bm\pi^2-e\Sigma_3B}.
\end{array} \label{THaem} \end{equation} We suppose that the energy is measured in electron volts.

If and only if the electric field is not strong ($|e\bm E|\ll\epsilon^2$), we can neglect the spin term. In this case, the operator equations of motion in the FW representation are given by
\begin{equation}\begin{array}{c}
\bm v\equiv \frac{d\bm r}{dt}=\frac i\hbar[{\cal H}_{FW},\bm r]=\beta v_F^2\frac{\bm\pi}{\epsilon},\\ \bm F\equiv \frac{d\bm\pi}{dt}=\frac i\hbar[{\cal H}_{FW},\bm\pi]=e\bm E+e\beta v_F^2\frac{\bm\pi}{\epsilon}\times\bm B\\=e(\bm E+\bm v\times\bm B).
\end{array} \label{EoM} \end{equation}
To clear the consideration, we have written the equation for the force $\bm F$ in the (3+1) laboratory frame.

The absolute value of the electron velocity is almost equal to $v_F$. When $E$ is significantly less than $v_FB$, the energy spectrum remains discrete. However, the wave eigenfunctions are distorted and the energy levels are shifted. The stationary states remain multiwave because they are formed by infinite continuums of partial waves. When $E/(v_FB)>1$, the average velocity along the $y$ axis becomes nonzero and electrons ($e=-|e|$) move antiparallel to this axis. In this case, there are not discrete energy levels and the energy spectrum collapses. 

The presented consideration cannot replace a detailed quantum-mechanical analysis. However, it shows an applicability of the relativistic FW transformation to the description of a graphene electron in crossed uniform electric and magnetic fields.

\section{Multiwave Hermite-Gauss states and extraordinary 
properties of graphene electrons in the free space}\label{general}

Fortunately, multiwave HG states of graphene electrons in the free (2+1)-space can also be deduced from the corresponding HG states of particles in the free (3+1)-space. The latter states are well known. They are defined by the wave functions localized in $x$ and $y$ dimensions (see Refs. \cite{Kogelnik,Siegman,Pampaloni}): $\Phi_{FW}=\psi_n(x,z)\psi_m(y,z)$. These functions are derived in the paraxial approximation ($|p_z|>>|\bm p_\bot|$) and are the solutions of the paraxial equation. In the relativistic QM in the FW representation, this equation has the form \cite{photonPRA}
\begin{equation}
\begin{array}{c}
\left(\nabla^2_\bot+2ik\frac{\partial}{\partial
z}\right)\!\Phi_{FW}=0,\quad
\nabla^2_\bot=
\frac{\partial^2}{\partial x^2}+\frac{\partial^2}{\partial y^2},
\end{array}
\label{eqp}
\end{equation} where $\hbar k=p=\sqrt{\bm p_\bot^2+p_z^2}$ and $\hbar$ is omitted. {Equation (\ref{eqp}) has been obtained for (3+1)-space.} In this equation, an appropriate shift of the squared particle momentum has been made \cite{paraxialLandau,arXiv}. It is similar to a shift of the zero energy level in Schr\"{o}dinger QM.

Wave functions in optics are similar to FW ones. The functions $\psi_n(x,z)$ and $\psi_m(y,z)$ are normalized to unit: $$\int{\psi_{n}^\dag(x,z)\psi_{n}(x,z)dx}=1,\quad \int{\psi_{m}^\dag(y,z)\psi_{m}(y,z)dy}=1.$$ 
However, a HG wave in the free (3+1)-space can be localized in $x$ direction and unlocalized in $y$ one. Such a beam can also exist in the free (2+1)-space. In this space, the paraxial equation takes the form
\begin{equation}
\begin{array}{c}
\left(\frac{\partial^2}{\partial x^2}+2ik\frac{\partial}{\partial
y}\right)\Phi_{FW}=0.
\end{array}
\label{eqpto}
\end{equation}
The corresponding wave function reads
\begin{equation}
\begin{array}{c}
\Phi_{FW}=\exp{(ip_yy)}\psi_{N}(x,y)\eta,\\ \int{\psi_N^\dag(x,y)\psi_N(x,y) dx}=1,\\
\psi_N(x,y)=C_{N}H_N\left[\frac{\sqrt2x}{w(y)}\right]\exp{\biggl[-\frac{x^2}{w^2(y)}-ik\frac{x^2}{2R(y)}}\\{+i\left(N+\frac12\right)\Phi_G(y)\biggr]},\quad
C_{N}=\frac{2^{1/4}}{\pi^{1/4}w^{1/2}(y)\sqrt{2^{N}N!}},\\
w(y)=w_0\sqrt{1+\frac{4y^2}{k^2w_0^4}},\quad R(y)=y+\frac{k^2w_0^4}{4y},\\
\Phi_G(y)=\arctan{\left[\frac{kw^2(y)}{2R(y)}\right]}=\arctan{\left(\frac{2y}{kw_0^2}\right)},
\end{array}
\label{eq33new}
\end{equation}
where $k$ is the wave number of the beam, $w_0$ is the minimum beam
width, $R(y)$ is the radius of curvature of the wave front, and $\Phi_G(y)$ is the Gouy phase.

Equation 
(\ref{eq33new}) describes multiwave (structured) HG beams being infinite continuums of partial de Broglie waves in the free (2+1)-space. These beams do not spread. As well as the corresponding beams in the free (3+1)-space \cite{photonPRA}, such beams possess extraordinary properties. First of all, current carriers (electrons and holes) in multiwave states move slower than the same carriers in structureless ones. QM gives the simple explanation of the similar effect in the free (3+1)-space \cite{photonPRA}. 

All beam parameters are defined
by expectation values or eigenvalues of related operators. In the considered case, the group velocity operator depends on a hidden motion along the $x$ axis. As follows from Eq. (\ref{eHamf}) at $B=0$,
\begin{equation}
\begin{array}{c}
\bm v=\frac{\partial{\cal H}_{FW}}{\partial\bm p}=\frac{v_F\bm p}{p},\quad v=v_F.
\end{array}
\label{velocity}
\end{equation} We use the term ``hidden motion'' for a motion which does not contribute to expectation values of operators defining some components of the velocity and momentum but affects expectation values of squares of both these operators and eigenvalues of the energy operator. In the considered case, $\langle v_x\rangle=0,~\langle v_y\rangle\neq0$. However, the absolute value of the total momentum, $p$, is equal to $\mathfrak{E}/v_F$ and is the same for all partial waves. The paraxial approximation leads to $p_y=\sqrt{p^2-p_x^2}\approx p\left(1-\frac{p_x^2}{2p^2}\right)$. Thus, the beam moves as a whole with the averaged group velocity (cf. Ref. \cite{photonPRA}) 
\begin{equation}
\begin{array}{c}
\langle\bm v\rangle=v^2_F\frac{\langle p_y\rangle \bm e_y}{\mathfrak{E}}\approx v_F\bm e_y\left(1-\frac{v^2_F\langle p_x^2\rangle}{2\mathfrak{E}^2}\right).
\end{array}
\label{Lenfree}
\end{equation}
This effect is similar to the subluminality of structured light (see Ref. \cite{photonPRA} and references therein).

Graphene electrons in the HG states also acquire nonzero effective masses dependent on a quantum number. In the laboratory
frame, a graphene electron can be considered as a centroid with the constant laboratory
frame energy $\mathfrak{E}$ and the velocity $|\langle\bm v\rangle|<v_F$ defined by Eq. (\ref{Lenfree}). Certainly, the centroid has also the nonzero effective mass (cf. Ref. \cite{photonPRA}) 
\begin{equation}
M=\frac{\sqrt{\mathfrak{E}^2-v_F^2\langle p_y^2\rangle}}{v_F^2}=\frac{\langle p_x^2\rangle^{1/2}}{v_F}.
\label{Effmass}
\end{equation}

{In principle, these unusual effects can be observed. Certainly, this is a rather nontrivial task. For a generation of HG beams in (2+1)-space, one can adapt existing methods developed for a generation of HG beams of photons and electrons in (3+1)-space \cite{ChenHuang,Kong,GGuzzinati}. As follows from Eq. (\ref{Lenfree}), the difference $\langle\bm v\rangle-v_F\bm e_y$ strongly depends on the electron energy. Therefore, this difference can be substantially increased if the HG beam is decelerated by a (quasi)unuform electric field collinear to the $y$ axis. This field does not lead to a beam spread (see Sec. \ref{Electric}). We suppose that the decrease of the beam velocity by 5-10\% makes the effect observable.}

\section{Multiwave states of graphene electrons in a static electric 
field}\label{Electric}

The following analysis shows that graphene electrons in a static electric field can be in multiwave states. We can suppose that the static and nonuniform electric field is negligible far from the field source and a graphene electron is in a multiwave HG state in this area. Such a graphene beam is accelerated or decelerated by the electric field. Far from the field source, the potential energy vanishes. Importantly, the \emph{total} energy of each partial beam remains unchanged. As a result, the beam remains coherent and non-spreading. If the potential energy in any fixed point with the radius vector $\bm R_0$ is equal to $U_0=eA_0(\bm R_0)$, it becomes equal to $U=eA_0(\bm R)=e[A_0(\bm R_0)-\bm E(\bm R_0)\cdot\bm r]$ ($\bm r=\bm R-\bm R_0$) in a nearby point with the radius vector $\bm R$. The simplest situation takes place when the electric field is collinear to the beam direction ($y$ axis). In this case, $p_x$ is not affected by the electric field. The paraxial approximation results in
\begin{equation}
\begin{array}{c}
\epsilon=p=\sqrt{p_x^2+p_y^2}\approx p_y+\frac{p_x^2}{2p}.
\end{array}
\label{mwave}
\end{equation} For graphene electrons, the Fermi velocity is equivalent to $c$ for photons and is here omitted. We suppose that $p_y>0$ and $\bm r=(x,y)$. Since $p_x$ is small, the last term in Eq. (\ref{THamf}) can be neglected. As a result, the energy in positive-energy states is given by 
\begin{equation}
\begin{array}{c}
\mathfrak{E}=p+U_0-E_yy.
\end{array}
\label{energ}
\end{equation}
When $y=0$, $p=p_0=\mathfrak{E}-U_0$. In the general case, $p=p_0+E_yy$. Therefore, Eq. (\ref{mwave}) takes the form
\begin{equation}
\begin{array}{c}
2(p_0+E_yy)^2\approx 2(p_0+E_yy)p_y+p_x^2.
\end{array}
\label{mwavenf}
\end{equation}
Since $E_y=const$, the operator expressions $yp_y$ and $p_yy$ [$p_y=-i\partial/(\partial y)$] are equivalent for the next transformation. This equation can be presented in the form
\begin{equation}
\begin{array}{c}
p_x^2+2(p_0+E_yy)p_y-4p_0E_yy-2E_y^2y^2\approx 2p_0^2.
\end{array}
\label{mwavefl}
\end{equation}
When we denote $k_0=p_0/\hbar$ and omit $\hbar$, the operator form of Eq. (\ref{mwavefl}) reads 
\begin{equation}
\begin{array}{c}
\left[-\frac{\partial^2}{\partial x^2}-2i(k_0+E_yy)\frac{\partial}{\partial y}-2(2k_0+E_yy)E_yy\right]\Phi_{FW}\\ \approx 2k_0^2\Phi_{FW}.
\end{array}
\label{mwavefo}
\end{equation} The substitution $\Phi_{FW}= \exp{(ik_0y)}\Psi$ brings the corresponding
paraxial equation
\begin{equation}
\begin{array}{c}
\left[\frac{\partial^2}{\partial x^2}+2i(k_0+E_yy)\frac{\partial}{\partial y}+2(2k_0+E_yy)E_yy\right]\Psi=0.
\end{array}
\label{pareq}
\end{equation}

A comparative analysis of Eqs. (\ref{eqpto}) and (\ref{pareq}) shows that wave functions of graphene electrons in the free (2+1)-space and a static electric field are similar and have the form (\ref{eq33new}) in both cases. However, in the latter case the parameters $w(y),\,R(y)$, and $\Phi_G(y)$ are not defined by Eq. (\ref{eq33new}) and differently depend on $y$.

Thus, graphene electrons in a static electric field can be in multiwave states which characterize non-spreading coherent beams. As a result, a static electric field can accelerate and decelerate HG beams of graphene electrons.
 
\section{Summary}\label{Discussionsummary}

The relativistic FW transformation has been used for an advanced description of free and interacting planar graphene electrons. It has been proven that the initial Dirac equation in (2+1)-space should be based on $4\times4$ Dirac matrices but not on {the reduction of matrix dimensions and the use of} $2\times2$ Pauli matrices. {The latter approach does not agree with the experiment.} New properties of graphene electrons in free (2+1)-space and external fields have been found. The exact 
FW Hamiltonian of a graphene electron in uniform and nonuniform magnetic field has been derived. It has been shown that spin effects are rather important and the spin takes the values $\pm1/2$. The exact energy spectrum agreeing with experimental data \cite{GrapheneMiller,GraphenePCheng} and exact FW wave eigenfunctions have been obtained. These eigenfunctions describe multiwave states in (2+1)-space. It has been proven that HG beams exist even in the free space. Two extraordinary 
properties of graphene electrons in the HG states are acquiring nonzero effective masses dependent on a quantum number and moving with group velocities which are less than the carrier velocity. In a static electric field, graphene electrons also can be in multiwave HG states defining non-spreading coherent beams. This means that HG beams of graphene electrons can be accelerated and decelerated.

\smallbreak

\textbf{Acknowledgments.}
{The author is grateful to Y.N. Obukhov and anonymous referees for valuable comments.} The author {also} acknowledges the support by the Chinese Academy of
Sciences President’s International Fellowship Initiative (Grant No. 2019VMA0019) and
hospitality and support by the Institute of Modern Physics of the Chinese Academy of
Sciences.


\begin{thebibliography}{}

\bibitem{GeimNov}
A. K. Geim, K. S. Novoselov, The rise of graphene,
Nature Materials \textbf{6}, 183 
(2007).

\bibitem{JMP}
A.\,J. Silenko, Foldy-Wouthuysen transformation for
relativistic particles in external fields, J. Math. Phys. {\bf 44}, 2952 (2003).

\bibitem{PRA2015}
A.\,J. Silenko, General method of the relativistic Foldy-Wouthuysen transformation
and proof of validity of the Foldy-Wouthuysen Hamiltonian, Phys. Rev. A \textbf{91}, 022103 (2015).

\bibitem{PRAnonstat}
A. J. Silenko, Energy expectation values of a particle in nonstationary fields, 
Phys. Rev. A \textbf{91}, 012111 (2015).

\bibitem{PRAFW}
Liping Zou, Pengming Zhang, and A. J. Silenko, Position and spin in relativistic quantum mechanics, Phys. Rev. A \textbf{101}, 032117 (2020). 

\bibitem{Landau}
L. D. Landau, Diamagnetismus der Metalle, Z. Phys. \textbf{64}, 629 (1930). 

\bibitem{LL3}
L. D. Landau, E. M. Lifshitz, \emph{Quantum Mechanics. Non-Relativistic Theory}, 3rd ed.
(Pergamon Press, Oxford, 1977), pp. 458-461.

\bibitem{GrapheneMiller}
D. L. Miller, K. D. Kubista, G. M. Rutter, M. Ruan, W. A. de Heer, P. N. First, J. A. Stroscio, Observing the Quantization of Zero Mass Carriers Observing the Quantization of Zero Mass Carriers in Graphene, Science \textbf{324}, 924 (2009).

\bibitem{GraphenePCheng}
P. Cheng \emph{et al.}, 
Landau Quantization of Topological Surface States in Bi$_2$Se$_3$, Phys. Rev. Lett. \textbf{105}, 076801 (2010).

\bibitem{Jiang}
Yeping Jiang, Yilin Wang, Mu Chen, Zhi Li, Canli Song, Ke He, Lili Wang, Xi Chen, Xucun Ma, and Qi-Kun Xue, Landau Quantization and the Thickness Limit of Topological Insulator Thin Films of Sb$_2$Te$_3$, Phys. Rev. Lett. \textbf{108}, 016401 (2012).

\bibitem{GrapheneGeim}
A. K. Geim, Graphene: Status and Prospects, Science \textbf{324}, 1530 (2009).

\bibitem{FW}
 L.\,L. Foldy, S.\,A. Wouthuysen, On the Dirac Theory of Spin 1/2
Particles and Its Non-Relativistic Limit,
Phys. Rev. \textbf{78}, 29 (1950).

\bibitem{Graphenesimulations}
G. Murgu\'{i}a1, A. Raya and \'{A}. S\'{a}nchez, Planar Dirac Fermions in External
Electromagnetic Fields, in \emph{Graphene Simulation},
ed. by Jian Ru Gong (IntechOpen, London, 2011).

\bibitem{JINRLett12}
A. J. Silenko, Classical limit of relativistic quantum mechanical equations in the Foldy-Wouthuysen representation,
Pis'ma Zh. Fiz. Elem. Chast. Atom. Yadra \textbf{10},
144 (2013) [Phys. Part. Nucl. Lett. \textbf{10}, 91 (2013)].

\bibitem{PRA2008}
A.\,J. Silenko, Foldy-Wouthuysen transformation and semiclassical limit for relativistic particles
in strong external fields, Phys. Rev. A \textbf{77}, 012116 (2008).

\bibitem{Unal}
Y. Sucu, N. \"{U}nal, Exact solution of Dirac equation in 2+1 dimensional
gravity, Found. Phys. \textbf{48}, 
052503 (2007).

\bibitem{Koke}
C. Koke, C. Noh, and D. G. Angelakis, Dirac equation on a square waveguide lattice with site-dependent coupling strengths and the gravitational Aharonov-Bohm effect, Phys. Rev. D \textbf{102}, 013514 (2020).

\bibitem{RJackiw}
R. Jackiw, Fractional charge and zero modes for planar systems in a magnetic field, Phys. Rev. D \textbf{29}, 2375 (1984).

\bibitem{Lozovik}
Y. E. Lozovik, S. P. Merkulova and A. A. Sokolik, Collective electron phenomena in graphene, Phys.-Usp. \textbf{51}, 727 (2008).

\bibitem{BLP}
V. B. Berestetskii, E. M. Lifshitz, and L. P. Pitayevskii,
{\em Quantum Electrodynamics}, 2nd ed. (Pergamon, Oxford, 1982).

\bibitem{PhysRevD2014}
A. J. Silenko and O. V. Teryaev, Spin effects and compactification, Phys. Rev. D \textbf{89}, 041501(R) (2014).

\bibitem{Horvathy}
P. A. Horv\'{a}thy, L. Martina, P. Stichel, Comments on spin-orbit interaction of anyons, Mod. Phys. Lett. A \textbf{20}, 1177 (2005). 

\bibitem{Case} K. M. Case, Some Generalizations of the Foldy-Wouthuysen Transformation,
Phys. Rev. \textbf{95}, 1323 
(1954).

\bibitem{Energy1}
W. Tsai, Energy eigenvalues for charged particles in a homogeneous
magnetic field -- an application of the Foldy-Wouthuysen transformation,
Phys. Rev. D \textbf{7},
1945 (1973).

\bibitem{Energy3}
A. J. Silenko, Connection between wave functions in the Dirac and Foldy-Wouthuysen
representations, Phys. Part. Nucl. Lett. \textbf{5}, 501 (2008).

\bibitem{spin1}
A. J. Silenko, High precision description and new properties of a spin-1 particle in a magnetic field, Phys. Rev. D \textbf{89},
121701(R) (2014).

\bibitem{paraxialLandau}
Liping Zou, Pengming Zhang, and A. J. Silenko, Paraxial wave function and Gouy phase for a relativistic electron in a uniform magnetic field, J. Phys. G: Nucl. Part. Phys. \textbf{47}, 055003 (2020).

\bibitem{Energy2}
A. A. Sokolov and I. M. Ternov, \textit{Radiation from
relativistic electrons}, 2nd ed. (AIP, New York, 1986).

\bibitem{Rajabi}
A. Rajabi and J. Berakdar, Relativistic electron vortex beams in a constant
magnetic field, Phys. Rev. A \textbf{95}, 063812 (2017).

\bibitem{CastroNeto}
A. H. Castro Neto, F. Guinea, N. M. R. Peres, K. S. Novoselov, and A. K. Geim, The electronic properties of graphene, Rev. Mod. Phys. \textbf{81}, 109 (2009).

\bibitem{GrapheneKuru}
\c{S}. Kuru, J. Negro and L. M. Nieto, Exact analytic solutions for a Dirac
electron moving in graphene under magnetic fields, J. Phys.: Condens. Matter \textbf{21}, 455305 (2009).

\bibitem{Haldane}
F. D. M. Haldane, Model for a Quantum Hall Effect without Landau Levels:
Condensed-Matter Realization of the ``Parity Anomaly'', Phys. Rev. Lett. \textbf{61}, 2015 (1988).

\bibitem{GraphenePeres}
N. M. R. Peres and E. V. Castro, Algebraic solution of a graphene layer in transverse
electric and perpendicular magnetic fields, J. Phys.: Condens. Matter \textbf{19}, 406231 (2007).

\bibitem{CastilloCeleita}
M. Castillo-Celeita1a, A. Contreras-Astorga and D. J. Fern\'{a}ndez C., Complex supersymmetry in graphene, Eur. Phys. J. Plus \textbf{137}, 904 (2022).

\bibitem{GrapheneNath}
D. Nath, M. Presilla, O. Panella and P. Roy, Non-commutativity effects in the Dirac equation in crossed electric and magnetic fields, EPL \textbf{123}, 20008 (2018).

\bibitem{VLukose}
V. Lukose, R. Shankar, and G. Baskaran, Novel Electric Field Effects on Landau Levels in Graphene, Phys. Rev. Lett. \textbf{98}, 116802 (2007).

\bibitem{Ma}
Ning Ma, Shengli Zhang, Daqing Liu, Erhu Zhang, Novel electric field effects on magnetic oscillations in graphene nanoribbons, Phys. Lett. A \textbf{375}, 3624 (2011). 

\bibitem{Betancur}
Y. Betancur-Ocampo, E. D\'{i}az-Bautista, and T. Stegmann, Valley-dependent time evolution of coherent electron states in tilted anisotropic Dirac materials, Phys. Rev. B \textbf{105}, 045401 (2022).

\bibitem{Ates}
\.{I}. B. Ate\c{s}, \c{S}. Kuru, J. Negro, Graphene Dirac fermions in symmetric electric and magnetic fields: the case of an electric square well, Phys. Scr. \textbf{98}, 015816 (2023).

\bibitem{Phan}
A.-L. Phan, D.-N. Le, V.-H. Le, P. Roy, Electronic spectrum in 2D Dirac materials under strain, Physica E \textbf{121}, 114084 (2020).

\bibitem{Le}
D.-N. Le, V.-H. Le, and P. Roy, Graphene under uniaxial inhomogeneous strain and an external electric field: Landau levels, electronic, magnetic and optical properties, Eur. Phys. J. B \textbf{93}, 158 (2020).

\bibitem{LeLe}
D.-N. Le, V.-H. Le, and P. Roy, Modulation of Landau levels and de Haas-van Alphen oscillation in magnetized graphene by uniaxial tensile strain/ stress, J. Magn. Magn. Mater. \textbf{522}, 167473 (2021).

\bibitem{Nimyi}
I. O. Nimyi, V. K\"{o}nye, S. G. Sharapov, and V. P. Gusynin, Landau level collapse in graphene in the presence of in-plane radial electric and perpendicular magnetic fields, Phys. Rev. B \textbf{106}, 085401 (2022).

\bibitem{Kogelnik}
H. Kogelnik and T. Li, Laser Beams and Resonators, Appl. Opt. \textbf{5}, 1550
(1966).

\bibitem{Siegman}
A. E. Siegman, \emph{Lasers} (University Science Books,
Sausalito, 1986).

\bibitem{Pampaloni}
F. Pampaloni, J. Enderlein, Gaussian, Hermite-Gaussian, and
Laguerre-Gaussian beams: A primer, arXiv:physics/0410021 (2004).

\bibitem{photonPRA}
A. J. Silenko, Pengming Zhang, and Liping Zou, Relativistic quantum-mechanical description of twisted paraxial electron and photon beams, Phys. Rev. A \textbf{100}, 030101(R) (2019).

\bibitem{arXiv}
Liping Zou, Pengming Zhang, and A. J. Silenko, General quantum-mechanical solution for twisted electrons in a uniform magnetic field, Phys. Rev. A \textbf{103}, L010201 (2021).

\bibitem{ChenHuang}
Y. F. Chen, T. M. Huang, C. F. Kao, C. L. Wang, S. C. Wang, Generation of Hermite-Gaussian modes in fiber-coupled laser-diode end-pumped lasers, IEEE J. Quantum Electron. \textbf{33}, 1025 (1997).

\bibitem{Kong} W. Kong, A. Sugita, and T. Taira, Generation of Hermite–Gaussian modes and vortex arrays based on two-dimensional gain distribution controlled microchip laser, Opt. Lett. \textbf{37}, 2661 (2012).

\bibitem{GGuzzinati}
G. Guzzinati, A. B\'{e}ch\'{e}, H. Louren\c{c}o-Martins, J. Martin, M. Kociak and J. Verbeeck, Probing the symmetry of the potential of localized surface plasmon resonances with phase-shaped electron beams, Nature Commun. \textbf{8}, 14999 (2017).

\end{thebibliography}
\end{document}